\documentstyle[eqsecnum,aps,preprint]{revtex}
\begin{document}\tightenlines

\draft
\title{Energy and momentum density of thermal gluon oscillations}
\author{H. Arthur Weldon}
\address{Department of Physics, West Virginia University,\\
Morgantown, WV 26506-6315, USA}

\maketitle

\def\pol{\vec}
\def\half{{1\over 2}}

\begin{abstract}
In the exact propagator for finite temperature gluons the  
location of the transverse
 and longitudinal poles in the
gluon propagator are unknown functions of wave vector: 
$\omega_{T}(k)$ and
$\omega_{L}(k)$. The residues of the poles, also unknown,  fix the
normalization of the 
one gluon vector potential and thus of the field strength. The naive
 energy density  $\pol{E}\cdot\pol{D}+\pol{B}\cdot\pol{H}$ is not
correct because of dispersion. By keeping the modulations due to 
the source currents
the energy density is shown to be  
 $\omega_{T}/V$ and $\omega_{L}/V$ regardless of the functional form of 
$\omega_{T}(k)$ and
$\omega_{L}(k)$.  The momentum density
is $k/V$. The resulting energy-momentum tensor is not symmetric. 
\end{abstract}


\section{ Introduction}

{\it Zero temperature.} At zero temperature the canonical operator 
formalism guarantees that a
 gluon of frequency $\omega$ has a field configuration with energy
density $\omega/V$.   The same  result
can be obtained from the  propagator because  the residue
of the pole in the  propagator at $\omega=k$ fixes the potential
 of a transverse gluon to be 
\begin{equation}
\pol{A}=\hat{\epsilon}\;\Big({Z_{3}\over
2kV}\Big)^{1/2}\;e^{i(\pol{k}\cdot \pol{r}-kt)}\end{equation}
The resulting electric and magnetic fields have an energy density
$(|\pol{E}|^{2}+|\pol{B}|^{2})/Z_{3}=\omega/V$.
In a covariant gauge the factor $1/Z_{3}$  may be
viewed as the constant value of the electric permittivity and magnetic
permeability $1/\epsilon=\mu=Z_{3}$ for on-shell gluons in the vacuum,
and the energy density is
$\epsilon|\pol{E}|^{2}+|\pol{B}|^{2}/\mu=\omega/V$. It is important to
realize that  $Z_{3}$ is a complicated function of the coupling
constant and depends on the infrared and ultraviolet regularization.
But the
statement that the energy density is $\omega/V$ is  exact. It
is useful to restate this result.   Suppose the full zero-temperature
effective action $\Gamma(A,\psi)$ were known and from this the
 energy-momentum tensor $T^{\mu\nu}(A,\psi)$ could be constructed. The energy
density $T^{00}$ would contain terms quadratic in $A$, as well as cubic 
 and higher powers  and also terms containing $\psi$. If the value of
$A$ is put at the one-gluon value (1.1) it is the quadratic term in the
exact $T^{00}$ that would equal $\omega/V$. Terms in $T^{00}$ that contain
higher powers of $A$ do not describe the energy of a single on-shell
gluon and should not be included.

{\it Hard thermal loop approximation.} At non-zero temperature
the analysis of the field energy of thermal gluon oscillations 
 has previously been done  within the hard thermal loop
approximation of Braaten and Pisarski [1-3]. This is a consistent
high-temperature  approximation to QCD in which only the order $T^{2}$
contributions of all one-loop diagrams are retained. The gluon
self-energy is gauge invariant and satisfies $K_{\mu}\Pi^{\mu\nu}_{\rm
htl}=0$. The poles  in the
gluon propagator are not on the light cone. For the two polarizations
transverse to the wave-vector $\pol{k}$ the pole is at a
frequency $\omega_{T}(k)$ that is proportional to $gT$ but with a
complicated dependence on $k$. There is another pole in the
propagator for gluons polarized longitudinally, i.e.  parallel to
$\pol{k}$, and this pole is at  a different frequency
$\omega_{L}(k)$. The energy density of the fields is  not
proportional to  $|\pol{E}|^{2}+|\pol{B}|^{2}$.
The analysis of the energy and momentum has been specific to the detailed
properties of the hard thermal loop approximation. 
Because the off-shell amplitudes with any number of external
 gluons have the remarkable property of
being  gauge-fixing independent and obeying abelian Ward 
identities [2,4], 
it is possible to  construct  effective actions that generate 
all the hard thermal
loops [3,5-11].  The 
effective actions are highly non-local and
consequently the canonical procedures for constructing the 
energy-momentum tensor do not apply. 
The full energy-momentum tensor contains
 quadratic and all higher
all powers of the vector potential $A^{\mu}$ since they summarize all 
higher correlations. From the Braaten-Pisarski form  of the action [3], 
I constructed a conserved energy
momentum tensor [12] but the quadratic part of that $T^{00}$  does not 
give the physical 
 energy density $\omega_{a}/V$ for the two modes.
Other authors have constructed  different conserved
 energy-momentum tensors [13-15], also containing all higher
powers of $A^{\mu}$, and have generalized them to curved space-time
[16]. 
The non-uniqueness of   $T^{\mu\nu}$ is expected since 
a term $\partial_{\alpha}X^{\alpha\mu\nu}$ will be conserved with
respect to $\partial_{\mu}$ for any $X^{\alpha\mu\nu}$ that is
antisymmetric under interchange of $\alpha$ and $\mu$.
Blaizot and Iancu [15]  found an energy-momentum
 tensor that does give the correct energy densities for the transverse
 and longitudinal  plane waves.
It differs from [12] by such a total derivative term.  
 In these analyses
 the detailed structure of the quadratic,
 cubic, and all higher correlations 
are related by the Ward identities. Consequently even in the 
quadratic approximation the specific $k$- dependence of $\omega_{T}(k)$
and $\omega_{L}(k)$  is essential.

The purpose of the present paper is to go beyond the hard thermal loop
approximation in computing the field energy of the
thermal gluon oscillations. This question is  simple enough 
that it can be answered exactly just from the structure of the exact gluon
propagator [17]. The
inverse of the propagator gives the quadratic term in effective action. 
As emphasized above, experience with the hard thermal loop approximation
shows that the spatial and temporal non-localities of the quadratic
action prevent a unique construction of even the quadratic terms in
energy-momentum tensor.  In this paper  a  physical argument will be
used along the following lines.  First, the quadratic terms in the action
are used to obtain the linear  field equation $\delta \Gamma(A)/\delta
A_{\mu}(x)=J^{\mu}(x)$  for the gluon potentials in the presence of an
arbitrary color current $J^{\mu}$. In coordinate space this will turn out
to be Maxwell's equation 
\begin{equation}
\partial_{\mu}H^{\mu\nu}=J^{\nu}_{\rm free}+{\cal O}(J^{2})
\label{max1}\end{equation} 
in a media characterized by an electric permittivity $\epsilon(K)$
and a magnetic permeability $\mu(K)$ that are related to the gluon
self-energy. 
Thus  even in the exact theory, the gluon oscillations have many
features in common with the
classical electrodynamics of a classical plasma.  Although the color
electric field $\vec{E}$  and the color magnetic
induction field $\vec{B}$ are produced by all the quarks
and gluons, the contribution  of the moving quarks and gluons
to the self-energy tensor  produces polarization and magnetization
currents which are incorporated   into 
the electric displacement field $D_{\ell}
=\epsilon(K)E_{\ell}(K)=H_{0\ell}$ and the
magnetic  field $H_{\ell}(K)=B_{\ell}(K)/\mu(K)
=\epsilon_{\ell mn}H_{mn}/2$.
The sources of these macroscopic fields is referred to as the free
current  $J^{\mu}_{\rm free}$. The label `free' does not mean
trivial, but rather that the polarization and magnetization currents
are not included in this current as that would be double counting.
In terms of Feynman diagrams $J^{\mu}_{\rm free}$ is
one-gluon-irreducible because the self-energy effects of the gluon
are contained in the 
electric permittivity $\epsilon(K)$ and the magnetic
 permeability $\mu(K)$.  

The next step is to construct the energy and momentum density of the
gluon fields. The energy and momentum of a source-free plane wave cannot
be simply identified. Instead it is necesary to examine field
configurations that are not completely source-free and thus are not
precisely plane waves. The energy and momentum of such modulated waves
 can be  determined from the currents that radiate those
fields. The essential physical step is to  realize that the  power
supplied to the fields and the  force  exerted on
the fields is given by the real part of 
 $J^{\alpha *}_{\rm free}F_{\alpha}^{\;\;\nu}$ for harmonic time
dependence. This is expressed in the identity
 \begin{equation}
\partial^{\mu}\big(H^{*}_{\mu\lambda}F^{\lambda\nu}\big)
+\half H^{*}_{\lambda\alpha}\big(\partial^{\nu}F^{\lambda\alpha}\big)
=J_{\alpha\;\rm free}^{*}F^{\alpha\nu}.\label{}\end{equation}
 Since the right hand side gives the rate of energy and momentum input to
the fields by the currents, the left hand side must describe  the energy
and momentum content of the field configurations. To use
(1.3) one must solve Maxwell's equations in the presence of a weak source
$J^{\alpha}_{\rm free}$ that supplies energy and momentum to
the fields. The source produces modulated fields of the form
$\exp(ikz-i\omega t)\,f(z,t)$ where $\pol{E}, \pol{B}, \pol{D}, \pol{H}$
all have different modulation functions $f$ because of the dispersion.
Keeping terms to first order in $\partial f/\partial t$ and $\partial
f/\partial z$ reduces the left side of (1.3) to a total derivative:
\begin{equation}
\partial_{\mu}T^{\mu\nu}= {\rm Re}\,(J_{\alpha\;\rm
free}^{*}F^{\alpha\nu})  \end{equation}
Since there is absorption in the plasma the energy and momentum of 
any traveling wave will not truly be constant in time. In applying (1.4)
it is necessary to neglect the damping or, equivalently, to neglect
the imaginary parts of $\epsilon(K)$ and $\mu(K)$. This is a physical limitation. 
Thus although the poles in the gluon propagator occur at complex values
of frequency, only the real parts $\omega_{T}(k)$ and $\omega_{L}(k)$
will be retained.  In the limit that
the source $J^{\alpha}_{\rm free}\to 0$ the modulated fields become plane
wave fields whose magnitude is completely fixed by the residues of the
poles in the propagator. For these plane wave modes of single gluons the
  energy-momentum tensor   in either mode is \begin{equation}
T^{\mu\nu}_{a}={1\over V}\left(\begin{array}{cc}
\omega_{a} & k^{n}\\
v_{a}^{m}\omega_{a} & v_{a}^{m}k^{n}\end{array}\right)
\hskip1cm a=T\;{\rm or}\; L
\end{equation}
where $v_{T}=d\omega_{T}/dk$ and $v_{L}=d\omega_{L}/dk$
are the group velocities. By the theorem of Kobes, Kunstatter and Rebhan [18]
the functions $\omega_{T}(k)$ and $\omega_{L}(k)$ are gauge invariant.
The components of
(1.4) satisfy simple physical checks. First,  the energy density
$T^{00}=\omega/V$ and the momentum density $T^{0n}=k^{n}/V$.
Because the energy and momentum are transported at the group velocity,
$T^{m0}=v^{m}T^{00}$ and $T^{mn}=v^{m}T^{0n}$.
No assumption is made about the form of the
dispersion relations $\omega_{T}(k)$ and $\omega_{L}(k)$.
(Note that $T^{\mu\nu}$ would be symmetric for dispersion relations of the form
form $\omega=(k^{2}+m^{2})^{1/2}$ because then $\pol{v}\,\omega=\pol{k}$.
That is certainly not the case here.)

The analysis performed here is intended to be  simple and 
physical. It is a generalization of old results in optics and classical
plasmas [19-21]. In those contexts  $\epsilon$ is assumed independent of
$k$ and $\mu=1$. Most treatments are rather unphysical in that they omit
discussion of the external current that supplies the energy and momentum
content of the fields even thought the modulated fields are only possible
with external sources.

The remainder of the paper is organized so that all physics is deduced 
from the exact thermal gluon propagator.
Sec 2 uses the  poles in the propagator to
obtain the  normalized transverse and longitudinal plane waves.
Sec 3 derives Maxwell's equations from the gluon propagator  and the
relation of $\epsilon(K)$ and $\mu(K)$ to the gluon self-energy. 
Sec 4 deduces the energy and momentum for the transverse mode. Sec 5 does
the same for the longitudinal mode.

\section{Poles in the thermal gluon propagator}

\subsection{Gluon propagator in general linear gauges}

One of the first obstacles in constructing the exact propagator
for finite temperature gluons is that $K_{\mu}\Pi^{\mu\nu}\neq 0$ [22-25]. 
It was shown in [17] that for any linear gauge-fixing condition the
Slavnov-Taylor implies that ${\rm Det}\big[-K^{2}\delta^{\mu}_{\;\;\nu}
+K^{\mu}K_{\nu}+\Pi^{\mu}_{\;\;\nu}\big]=0$. Local gauge
invariance requires  one of the four eigenvalues to be
zero for all $K^{\mu}$. This provides a non-linear
condition on the self-energy tensor [24-28]. It is easiest to work
with a gauge-fixing that preserves the rotational invariance of the
plasma rest frame. These include covariant gauges, Coulomb gauges, and
temporal gauges. In rotationally covariant gauges the  gluon
polarizations that are transverse to the spatial momentum $\vec{k}$
are isolated by the projection tensor
\begin{equation}
A^{\mu\nu}\Big|_{\rm rest}=\left(\begin{array}{cc}
0&0\\0&-\delta^{mn}+\hat{k}^{m}\hat{k}^{n}\end{array}
\right).\end{equation}
The gluon four-momentum is $K^{\mu}$ and a related
vector in which the energy and momentum components are reversed is
denoted by $\tilde{K}^{\mu}$:
\begin{equation}
K^{\mu}=(\omega, \vec{k})\hskip1cm
\tilde{K}^{\mu}=(k,\omega\hat{k}) \end{equation}
The new vector satisfies $\tilde{K}^{2}=-K^{2}$ and
$\tilde{K}\cdot K=0$. The same construction applies to any vector 
whose spatial component is along $\hat{k}$:
\begin{equation}
V^{\mu}=(v_{0},v\hat{k})\hskip1cm \tilde{V}^{\mu}
=(v,v_{0}\hat{k})\end{equation}
These satisfy $\tilde{V}^{2}=-V^{2}$ and $\tilde{V}\cdot V=0$.
For a linear gauge-fixing condition that preserves the rotational
invariance the inverse of the free propagator has the form
\begin{equation}
\Big[D^{-1}_{\rm free}(K)\Big]^{\mu\nu}
=-K^{2}g^{\mu\nu}+K^{\mu}K^{\nu}-{F^{\mu}F^{\nu}\over
\xi}\end{equation}
where the gauge-fixing vector $F^{\mu}$ has spatial components
parallel to $\hat{k}$.
 The free propagator  is 
\begin{equation}D^{\mu\nu}_{\rm free}(K)=
-{A^{\mu\nu}\over
K^{2}}+{\tilde{F}^{\mu}\tilde{F}^{\nu}\over (F\cdot
K)^{2}} -\xi {K^{\mu}K^{\nu}\over (F\cdot K)^{2}}																		
\end{equation}
Familiar examples of this include covariant gauges
$[ \tilde{F}^{\mu}=\tilde{K}^{\mu}]$, Coulomb gauges $[\tilde{F}^{\mu}
=(k,0,0,0)]$, and temporal gauges $[\tilde{F}^{\mu}=(0,\hat{k})]$.
Including interactions gives a proper self-energy tensor of the
form
 \begin{equation}\Pi^{\mu\nu}
=\Pi_{T}A^{\mu\nu}+\Pi_{L}{\tilde{K}^{\mu}\tilde{K}^{\nu}\over
\tilde{K}^{2}}
+\Pi_{C}{K^{\mu}\tilde{K}^{\nu}+\tilde{K}^{\mu}K^{\nu}\over K^{2}}
+\Pi_{D}{K^{\mu}K^{\nu}\over K^{2}}
\end{equation}
because of rotational invariance [24-28].  
(For QED and also for QCD in the hard thermal loop
approximation   $\Pi_{C}=\Pi_{D}=0$.)  The Slavnov-Taylor
condition leads to 
$\Pi_{C}=\sigma\sqrt{(K^{2}-\Pi_{L})\Pi_{D}}$.
and the full propagator is
\begin{equation}D^{\mu\nu}(K)=-{A^{\mu\nu}\over K^{2}-\Pi_{T}}
+K^{2}{\tilde{F}^{\mu}\tilde{F}^{\nu}\over (F\cdot H)^{2}}
-\xi {H^{\mu}H^{\nu}\over (F\cdot H)^{2}},
\label{prop}\end{equation}
where 
$ H^{\mu}=K^{\mu}\sqrt{K^{2}-\Pi_{L}}+\sigma\tilde{K}^{\mu}
\sqrt{\Pi_{D}}$. See also  [22,27].
The second denominator may be written directly in terms of the self-energy
tensor as
\begin{equation}
{(F\cdot H)^{2}\over K^{2}}=(F\cdot K)^{2}+\tilde{F}_{\mu}
\tilde{F}_{\nu}\Pi^{\mu\nu}
\end{equation}
The longitudinal denominator for three simple gauge choices are as follows:
\begin{eqnarray}
{\rm Covariant:}\hskip1cm  {(F\cdot H)^{2}\over
K^{2}}&=&(K^{2})^{2}+\tilde{K}_{\mu} \tilde{K}_{\nu}\Pi^{\mu\nu}\\
{\rm Coulomb:}\hskip1cm  {(F\cdot H)^{2}\over K^{2}} &=&
k^{2}\big(k^{2}+\Pi^{00}\big)\\ 
{\rm Temporal:}\hskip1cm  {(F\cdot
H)^{2}\over K^{2}} &=& k_{0}^{2}+\hat{k}^{m}\hat{k}^{n}\Pi^{mn}
\end{eqnarray}
It is  sometimes convenient to parametrize the
gauge-fixing vector as
$F^{\mu}=f_{1}\,K^{\mu}+f_{2}\,\tilde{K}^{\mu}$ 
so that
$F\cdot H=K^{2}\big(f_{1}\sqrt{K^{2}-\Pi_{L}}-\sigma
f_{2}\sqrt{\Pi_{D}}\big)$.

\subsection{Transverse plane waves}

The poles in the propagator (\ref{prop}) determine the freely
propagating plane wave solutions. For the transverse mode
the value of $\omega$ for which $K^{2}=\Pi_{T}$ defines the complex
energy $\omega_{T}^{c}=\omega_{T}-i\gamma_{T}$.
At this pole
\begin{equation}
\omega\to\omega_{T}^{c}:\hskip1cm
D^{mn}(K)\to {Z_{T}\over 2\omega_{T}^{c}(\omega-\omega_{T}^{c})}
\;(\delta^{mn}-\hat{k}^{m}\hat{k}^{n}),\end{equation}
where the wave function renormalization factor is defined by
\begin{equation}
{2\omega_{T}^{c}\over Z_{T}}={\partial\over\partial\omega}
(K^{2}-\Pi_{T})\bigg|_{\omega_{T}^{c}}.\label{Ztrans}\end{equation}
The residue of the pole fixes the normalization of the vector potential
for one transverse gluon to be
\begin{equation}
\vec{A}=-i\hat{\epsilon}\Big({Z_{T}\over
2\omega_{T}^{c}V}\Big)^{1/2}\; e^{i(\pol{k}\cdot\pol{r}
-\omega_{T}^{c}t)}\end{equation}
with $V$  the quantization volume. The phase factor $i$ has been
chosen for later convenience. As usual 
$\hat{\epsilon}$ is a unit vector satisfying
$\pol{k}\cdot\hat{\epsilon}=0$.
For definiteness  the direction of propagation will be chosen as the $z$
axis and the direction of polarization as the $x$ axis: 
\begin{equation}
A_{x}=-i\Big({Z_{T}\over
2\omega_{T}^{c}V}\Big)^{1/2}\; e^{i(kz
-\omega_{T}^{c}t)}\end{equation}
The corresponding fields are
\begin{equation}
E_{x}=\omega_{T}^{c}\Big({Z_{T}\over
2\omega_{T}^{c}V}\Big)^{1/2}\; e^{i(kz
-\omega_{T}^{c}t)}
\hskip1cm
B_{y}=k\Big({Z_{T}\over
2\omega_{T}^{c}V}\Big)^{1/2}\; e^{i(kz
-\omega_{T}^{c}t)}\label{EBtrans}\end{equation}
It will be shown later that these fields propagate freely in that
they satisfy Maxwell's equations with an electric permittivity $\epsilon$
and a magnetic permeability $\mu$ but with with no free current.

\subsection{Longitudinal plane waves}

When $(F\cdot H)^{2}=0$ the propagator (\ref{prop}) has a different pole
which will be called longitudinal because the resulting electric field
is along the spatial wave vector $\pol{k}$ and there is no magnetic
field. The value of $\omega$ at which the denominator $(F\cdot H)^{2}$
vanishes defines the complex energy $\omega_{L}^{c}$, which has real and
imaginary parts $\omega_{L}^{c}=\omega_{L}-i\gamma_{L}$.
The propagator at this pole behaves as
\begin{equation}
\omega\to \omega_{L}^{c}:\hskip1cm
D^{\mu\nu}(K)\to {Z_{L}
\over 2\omega_{L}^{c}(\omega-\omega_{L}^{c})}
{\tilde{F}^{\mu}\tilde{F}^{\nu}\over K^{2}}
\end{equation}
where the $\xi$ dependent tensor in (\ref{prop}) is omitted.
The  wave function renormalization factor for the longitudinal mode is
 \begin{equation}
{2\omega_{L}^{c}\over Z_{L}}={\partial\over\partial\omega}
\Big({F\cdot H\over K^{2}}\Big)^{2}\;
\Bigg|_{\omega_{L}^{c}}.\label{Zlong}\end{equation}
The coefficient of the pole fixes the normalization of the vector
potential for one longitudinal gluon to be
\begin{equation}
A^{\mu}=-i\Bigg({Z_{L}\over 2\omega_{L}^{c}V K_{L}^{2}}\Bigg)^{1/2}
\tilde{F}^{\mu}\;e^{-i(\pol{k}\cdot\pol{r}-\omega_{L}^{c}t)}\end{equation}
where the phase factor $-i$ has been chosen for later convenience.
Since 
\begin{equation}
\tilde{F}^{\mu}=f_{1}\tilde{K}^{\mu}+f_{2}K^{\mu}=-f_{1}{K^{2}\over
k}u^{\mu}+(f_{1}{\omega\over k} +f_{2})K^{\mu}\end{equation}
the potential is
\begin{equation}
A^{\mu}=i{f_{1}\over k}\Bigg({Z_{L}K_{L}^{2}\over 2\omega_{L}^{c}V
}\Bigg)^{1/2} u^{\mu}\;e^{-iK_{L}\cdot
x}+(\cdots)\;K^{\mu}.\end{equation} 
The omitted terms proportional to $K^{\mu}$
do not contribute to the  abelian field strength. Only the potential
along $u^{\mu}$ is  non-trivial and in the plasma rest frame the 
 only non-vanishing field is
\begin{equation}
E_{z}=f_{1}\Bigg({Z_{L}K_{L}^{2}\over 2\omega_{L}^{c}V
}\Bigg)^{1/2}\;e^{i(kz-\omega_{L}^{c}t)}.
\label{Elong}\end{equation}
Since there is no magnetic field,   the way in which this mode
transports momentum is particularly subtle. 

\section{Maxwell's equations from the gluon propagator}

The previous section displayed the plane waves that come from the
propagator poles. This section will show that the off-shell gluon
propagator implies Maxwell's equations in a media.

\subsection{Electric and magnetic phenomena}

{\it Current conservation.}
The gluon propagator determines the quadratic term in the generating
functional for one-particle-irreducible diagrams in the presence of a
colored source current $J^{a}_{\mu}$:
 \begin{equation}
W(J)={1\over 2}\int d^{4}xd^{4}y\;
J^{a}_{\mu}(x)D^{\mu\nu}(x-y)J^{a}_{\nu}(y)
+{\cal O}(J^{3}).\label{2.4.1}\end{equation}
All four components of $J_{\mu}$ are independent 
variables, not constrained by current conservation.
In order to obtain the equation of motion for the gauge potentials it
is necessary to impose color conservation, $D_{\mu}^{ac}J^{c\mu}=0$, where
\begin{equation}
D_{\mu}^{ac}\equiv\delta^{ac}\partial_{\mu}-gf^{abc}{\delta W\over
\delta J^{b\mu}(x)}\end{equation}
To impose the constraint it is convenient to introduce a Lagrange
multiplier function $\psi(x)$ and vary all four components of
$J_{\mu}$ independently in the functional
 \begin{equation}
W^{\prime}(J)=W(J)-{1\over 2}\int d^{4}x\;\psi(x)
(D^{\mu}J_{\mu})^{a}(D^{\nu}J_{\mu})^{a}\end{equation}
The resulting vector potential $A^{\mu}=\delta W^{\prime}/
\delta J_{\mu}$ is
\begin{equation}
A^{a\mu}(x)=\int d^{4}y\;D^{\mu\nu}(x-y)J^{a}_{\nu}(y)
+\psi(x)\partial^{\mu}\partial^{\nu}J^{a}_{\nu}(x)
+{\cal O}(J^{2})\end{equation}
In momentum space
\begin{equation}
A^{a\mu}(K)=\Big[D^{\mu\nu}(K)
-\Psi K^{\mu}K^{\nu}\Big]J^{a}_{\nu}(K)+{\cal O}(J^{2}),
\label{3.1.5}\end{equation}
where $\Psi\equiv \psi(-K)$. 
From now on the color indices will be omitted.
To solve for the Lagrange multiplier function,
contract both sides of the above with the vector $F_{\mu}$ and use the
fact that (\ref{prop}) satisfies $F_{\mu}D^{\mu\nu}=-\xi H^{\nu}J_{\nu}
/(F\cdot H)$ to get:
 \begin{equation}
F_{\mu}A^{\mu}(K)
=-{\xi\over F\cdot H}H^{\nu}J_{\nu}(K)
-\Psi\; (F\cdot K)\;K^{\nu}J_{\nu}(K)\end{equation}
Substituting this into (\ref{3.1.5}) and rearranging gives
\begin{equation}
A^{\prime\mu}(K)
=\Big[D^{\mu\nu}(K)+\xi {K^{\mu}H^{\nu}\over (F\cdot K)(F\cdot H)}
\Big]J_{\nu}(K),\label{3.1.7}\end{equation}
where
\begin{equation}
A^{\prime\mu}(K)\equiv \Big[g^{\mu\nu}-{K^{\mu}
F^{\nu}\over K\cdot F}\Big]A_{\nu}(K).\end{equation}
This potential  satisfies the gauge condition
$F_{\mu}A^{\prime\mu}(K)=0$.
In coordinate space (\ref{3.1.7}) is the integral equation for
$A^{\prime\mu}$.  By multiplying from the left
by the  tensor shown below, (\ref{3.1.7}) can be converted into the
differential equation
 \begin{equation}
\Bigl[(-K^{2}+\Pi_{T})A^{\alpha\mu}
+\Big({F\cdot H\over
f_{1}K^{2}}\Big)^{2}{\tilde{K}^{\alpha}\tilde{K}^{\mu}\over
K^{2}}
\Bigr]A_{\mu}^{\prime}(K)=J^{\alpha}_{\rm
free}(K),\label{2.4.3}\end{equation}
 where the free current is defined as
\begin{equation}
J^{\alpha}_{\rm
free}(K)\equiv J^{\alpha}(K)+\xi\Pi_{D}\,{\tilde{K}^{\alpha}
\tilde{K}^{\nu}J_{\nu}(K)\over (f_{1}K^{2})^{2}}
\label{2.4.4}\end{equation} 
and satisfies $K_{\alpha}J^{\alpha}_{\rm free}(K)=0$.
The transverse plane waves of Sec 2.2 at $K^{2}-\Pi_{T}=0$
 propagate with $J_{\rm free}=0$. Similarly
the longitudinal plane waves of Sec 2.3 at $(F\cdot H)^{2}=0$ 
propagate with $J_{\rm free}=0$. 
It is sometimes useful to have the inverse form of (\ref{2.4.3}):
\begin{equation}
A^{\prime}_{\mu}(K)=\Big[-{A_{\mu\nu}\over K^{2}-\Pi_{T}}
+K^{2}{\tilde{F}_{\mu}\tilde{F}_{\nu}\over (F\cdot H)^{2}}
\Big]J^{\nu}_{\rm free}(K)
\end{equation}

{\it Identification of $\epsilon$ and $\mu$.}
Equations (\ref{2.4.3}) are   the inhomogeneous  Maxwell
 equations in a media characterized
by an electric permittivity $\epsilon$ and a magnetic permeability 
$\mu$. To see this one needs to separate the electric and magnetic component of
$\tilde{K}^{\mu}A_{\mu}^{\prime}(K)$ by using
  \begin{equation}
\hat{k}^{\ell}\tilde{K}^{\mu}A_{\mu}^{\prime}(K)(K)
=i\pol{E}^{\ell}(K)+\omega
\bigl(\delta^{\ell
m}-\hat{k}^{\ell}\hat{k}^{m}\bigr)A^{\prime
m}(K),\label{2.4.5}\end{equation} where $\pol{E}(K)=-i\pol{k}
A^{\prime 0}(K)+i\omega \pol{A}^{\prime}(K)(K)$.
 The $\alpha=0$ component
of (\ref{2.4.3}) is
 \begin{equation}
\epsilon(k,\omega)\;i\pol{k}\cdot\pol{E}(K)=J^{0}_{\rm free},
\label{Gauss1}\end{equation}
with a gauge-dependent  electric permitivity given by 
\begin{equation}
\epsilon(k,\omega)={1\over K^{2}}\Big({F\cdot H\over
f_{1}K^{2}}\Big)^{2}=1+{\tilde{F}_{\mu}\tilde{F}_{\nu}
\Pi^{\mu\nu}\over (F\cdot K)^{2}}.\label{eps}
\end{equation} The electric displacement field is
$\pol{D}(K)=\epsilon\,\pol{E}(K)$. Next comes the Ampere-Maxwell equation,
which is contained in the spatial components of (\ref{2.4.3}):
  \begin{equation}
(K^{2}-\Pi_{T})(-\delta^{\ell m}+\hat{k}^{\ell}\hat{k}^{m})A^{\prime
m}(K) +\epsilon(k,\omega)\omega\hat{k}^{\ell}\tilde{K}^{\mu}A_{\mu}
^{\prime}(K)=
J^{\ell}_{\rm free}.\end{equation}
Using (\ref{2.4.5}) this becomes
\begin{equation}
{1\over\mu(k,\omega)}\;
i\pol{k}\times\pol{B}(K)+\epsilon(k,\omega)i\omega\pol{E}(K)
=\pol{J}^{\rm ext},\label{Ampere1}\end{equation}
where $\pol{B}(K)=i\pol{k}\times\pol{A}^{\prime}(K)$ is the magnetic induction field and
the gauge-dependent magnetic permeability is 
\begin{equation}
{1\over\mu(k,\omega)}={1\over k^{2}}
\bigl[-K^{2}+\Pi_{T}+\omega^{2}\epsilon(k,\omega)\bigr]\label{mu}.
\end{equation}
The magnetic field is $\pol{H}(K)=\pol{B}(K)/\mu$.

\subsection{Energy and Momentum Conservation}

In coordinate space the inhomogeneous equations (\ref{Gauss1}) and
(\ref{Ampere1}) take the familiar form
\begin{equation}\pol{\nabla}\cdot\pol{D}=J^{0}_{\rm free}\hskip1cm
\pol{\nabla}\times\pol{H}-{\partial\pol{D}\over\partial t}
=\pol{J}_{\rm free}.\label{inhomo}\end{equation}
The  two homogeneous  equations, $\pol{\nabla}\cdot\pol{B}=0$ and 
$\pol{\nabla}\times\pol{E}+\partial\pol{B}/
\partial t=0$, are  true for any $A^{\mu}$.
These results can be expressed in terms of the tensors
\begin{equation}
H^{\mu\nu}=\left(\begin{array}{rrrr} 0 & -D_{x} &-D_{y} & -D_{z}\\
D_{x} & 0& -H_{z} & H_{y}\\
D_{y} & H_{z} &0& -H_{x}\\
D_{z} & -H_{y} & H_{x} & 0 \end{array}\right)
\hskip0.5cm
F^{\mu\nu}=\left(\begin{array}{rrrr} 0 & -E_{x} &-E_{y} & -E_{z}\\
E_{x} & 0& -B_{z} & B_{y}\\
E_{y} & B_{z} &0& -B_{x}\\
E_{z} & -B_{y} & B_{x} & 0 \end{array}\right).\label{HF}\end{equation}
Maxwell's equations in the rest frame of the plasma can be written
\begin{equation}
\partial_{\mu}H^{\mu\nu}=J^{\nu}_{\rm free}
\hskip2cm \partial^{\alpha}F^{\beta\gamma}
+\partial^{\beta}F^{\gamma\alpha}+\partial^{\gamma}
F^{\alpha\beta}=0.\label{cov}\end{equation}
In order to identify the energy and momentum in the fields, the
essential physical step is to recognize that the work done on  the
fields by the currents and the force exerted on the fields by the
currents  are given by 
the real part of $J^{*}_{\alpha\;\rm free}F^{\alpha\nu}$ for 
harmonic time dependence.  
This is  expressed in the identity 
\begin{equation}
\partial^{\mu}\big(H^{*}_{\mu\lambda}F^{\lambda\nu}\big)
+\half H^{*}_{\lambda\alpha}\big(\partial^{\nu}F^{\lambda\alpha}\big)
=J_{\alpha\;\rm free}^{*}F^{\alpha\nu},\label{cons}\end{equation}
which follows directly from (\ref{cov}).
The $\nu=0$ component is
\begin{equation}
-\pol{\nabla}\cdot(\pol{H}^{*}\times\pol{E})
+{\partial\pol{D}^{*}\over\partial t}\cdot\pol{E}
+\pol{H}^{*}\cdot{\partial \pol{B}\over\partial t}
=-\pol{J}^{*}_{\rm free}\cdot\pol{E}.\label{eng1}\end{equation}
The right hand side is the power supplied to
  the fields by the color charges 
in the source current. The left hand side contains the spatial
and temporal changes in the field energy as will later be shown.
The $\nu=1,2,3$ components of (\ref{cons}) are
\begin{eqnarray}
{\partial\over\partial t}(\pol{D}^{*}\times\pol{B})
+(\pol{\nabla}H^{*}_{j})B_{j}-&B_{j}\nabla_{j}\pol{H}^{*}
+D_{j}^{*}\pol{\nabla}E_{j}\nonumber\\
=-&J^{0*}_{\rm free}\pol{E}-\pol{J}^{*}_{\rm free}
\times\pol{B}.\label{mom1}\end{eqnarray}
The right hand side is the force acting on 
the fields by the color charges.
This external force is responsible for changes in the field momentum that
are contained on the left hand side.

\section{Transverse gluon oscillations}

\subsection{Modulated fields radiated from a current}

The transverse plane waves (\ref{EBtrans}) that correspond to poles
 in the propagator at $K^{2}=\Pi_{T}$
are exact solutions of Maxwell's equations (\ref{2.4.3}) with 
$J^{\mu}_{\rm free}=0$. 
As discussed in Sec 1, to identify the energy and momentum it is 
necessary to consider a modulated wave  of the form
 \begin{equation}
E_{x}=e^{i(kz-\omega_{T} t)}\;E_{T}(z,t),\label{3.1}
\end{equation}
where the amplitude function $E_{T}(z,t)$ is almost  constant.
The only significant variation in the amplitude function should be
when the wave is emitted or absorbed (i.e.  
at large $|z|$ and large $ |t|$). In Fourier space the amplitude
is \begin{equation} E_{T}(z,t)=\int d\alpha d\beta\; C_{T}(\alpha,\beta)
\;e^{i(\alpha z-\beta t)},\label{E0T}\end{equation}
where $C_{T}(\alpha,\beta)$ should have support only for very small wave
vector $\alpha$ and very small frequency $\beta$.  Note  that
this is not  a superposition of source-free fields, which would be
 the
case if $k+\alpha$  produced a pole at $\omega_{T}+\beta$.
Here $\alpha$ and $\beta$ are independent variables; the wave is 
slightly off the mass-shell.
The total wave vector and total frequency in (\ref{3.1}) will be
denoted 
\begin{equation}
k^{\prime}\equiv k+\alpha\hskip2cm
\omega^{\prime}\equiv\omega_{T}+\beta.\end{equation}
Faraday's law requires that the magnetic induction field be
\begin{equation}
B_{y}=e^{i(kz-\omega_{T} t)}\int d\alpha d\beta\;
{k'\over \omega'}C_{T}(\alpha,\beta) \;e^{i(\alpha z-\beta t)}.
\end{equation}
The constituitive relations  then fix the electric displacement
field and the magnetic field to be
\begin{equation}
D_{x}=e^{i(kz-\omega_{T} t)}\int d\alpha d\beta\;
\epsilon(k',\omega')C_{T}(\alpha,\beta) \;e^{i(\alpha z-\beta t)}
\end{equation}
\begin{equation}
H_{y}=e^{i(kz-\omega_{T} t)}\int d\alpha d\beta\;
{k'\over \omega'\mu(k^{\prime},\omega')}C_{T}(\alpha,\beta) 
\;e^{i(\alpha z-\beta
t)}. \end{equation}
The current which radiates these  modulated fields  can be easily
 computed from (\ref{inhomo}). The only non-vanishing component of
$J^{\mu}_{\rm free}$ is $J_{x}$ with
\begin{equation}
J_{x\;\rm free}=e^{i(kz-\omega_{T}t)}\int\hskip-0.2cm  d\alpha
d\beta\;\Big[ {k^{\,\prime
2}\over\omega^{\prime}\mu(k',\omega')}-\omega'
\epsilon(k',\omega')\Bigr] C_{T}(\alpha,\beta) \;e^{i(\alpha z-\beta
t)}. \end{equation}
 In the plane wave limit  
$C_{T}(\alpha,\beta)\to \delta(\alpha)\delta(\beta)E_{0}$ and 
the source will vanish, i.e.
$J_{x\;\rm free}\to 0$.

The fields $E_{x}, D_{x}, B_{y}, H_{y}$ each have slightly 
different modulations
multiplying the $\exp(ikz-i\omega_{T}t)$ factor. It is these slight
differences that will account for the energy balance. 
For example, in the integrand of  $B_{y}$ the factor 
$k'/\omega'$ can be expanded to first order as
\begin{equation}
B_{y}=e^{i(kz-\omega_{T} t)}\int d\alpha d\beta\;
\bigl({k\over\omega_{T}}+{1\over\omega_{T}}\alpha
-{k\over\omega_{T}^{2}}\beta+\cdots\bigr)C_{T}(\alpha,\beta) 
\;e^{i(\alpha
z-\beta t)}.\end{equation}
Only these terms are important since $C_{T}(\alpha,\beta)$ has
 support only
for very small values of $\alpha$ and $\beta$. Using (\ref{E0T}) 
this can be
written as
\begin{equation}
B_{y}=e^{i(kz-\omega_{T} t)}
\Bigl({k\over\omega_{T}}E_{T}-{i\over\omega_{T}}{\partial E_{T}
\over\partial z}
-{ik\over\omega_{T}^{2}}{\partial E_{T}\over \partial
t}+\cdots\Bigr).\label{B}\end{equation}
In the same fashion
\begin{equation}
D_{x}=e^{i(kz-\omega_{T}t)}\Bigl(\epsilon E_{T}-i{\partial\epsilon
\over\partial k}{\partial E_{T}\over\partial z}
+i{\partial \epsilon\over\partial\omega}
{\partial E_{T}\over\partial t}+\cdots\Bigr)\label{D}\end{equation}
\begin{equation}
H_{y}=e^{i(kz-\omega_{T}t)}\Bigl({k\over\omega_{T}\mu}E_{T}
-i{\partial \over\partial k}\Bigl({k\over\omega \mu}\Bigr)
{\partial E_{T}\over\partial z}
+i{\partial\over\partial\omega}\Bigl({k\over\omega\mu}\Bigr)
{\partial E_{T}\over\partial t}+\cdots\Bigr).\label{H}\end{equation}
Note that $\epsilon, \mu$, and the partial derivatives
 are to be evaluated at $\omega=\omega_{T}$. If $\epsilon$ and 
$\mu$ were
constants, then the above relations would reduce to
$D_{x}=\epsilon E_{x}$ and $H_{y}=B_{y}/\mu$. 

\subsection{Energy Conservation}

For the TEM mode with $\pol{E}$ polarized along the $x$ axis,
 the work-energy theorem (\ref{eng1})
is \begin{equation}
{\partial\over\partial z}{\rm Re}(H_{y}^{*}E_{x})
+{\rm Re}({\partial D_{x}^{*}\over\partial t}E_{x})
+{\rm Re}(H_{y}^{*}{\partial B_{y}\over\partial t}) 
=-{\rm Re}(J_{x\;\rm free}^{*}E_{x})\label{TPoyn2}\end{equation}
As discussed in Sec 1, it is necessary to treat $\epsilon$ and $\mu$ as
real in order to ultimately obtain a time-independent energy. For harmonic
fields and currents the real part of each product gives  one-half the
instantaneous value as will be seen below.
 To first order in the derivatives of $E_{T}$, the spatial divergence
term  gives\begin{equation}
{\partial\over\partial z}{\rm Re}(H_{y}^{*}E_{x})
={k\over\mu\omega_{T}}
\;{\partial\over\partial z}|E_{T}|^{2},\label{we1}\end{equation}
as expected. However the time derivatives in (\ref{TPoyn2}) become
\begin{equation}
{\rm Re}({\partial D_{x}^{*}\over\partial t}E_{x})
={\partial \over\partial\omega}(\omega\epsilon)\;
{\partial\over\partial t}{|E_{T}|^{2}\over 2}\
-\omega{\partial \epsilon\over\partial k}
{\partial\over\partial z}{|E_{T}|^{2}\over 2}\label{we2}\end{equation}
\begin{equation}
{\rm Re}(H_{y}^{*}{\partial B_{y}\over\partial t}) 
=-{\partial\over\partial \omega}\Bigl({k^{2}\over\mu\omega}\Bigr)
{\partial\over\partial t}{|E_{T}|^{2}\over 2}
+{k^{2}\over\omega_{T}}\Bigl({\partial\over\partial k}{1\over\mu}
\Bigr)
{\partial\over\partial z}{|E_{T}|^{2}\over 2}\label{we3}.\end{equation}
It is because of spatial dispersion, specifically because 
$\epsilon$ and $\mu$ depend on $k$,
that the right hand sides of (\ref{we2}) and (\ref{we3}) contain
 spatial derivatives of
$|E_{T}|^{2}$. The sum of  (\ref{we1})-(\ref{we3}) is
\begin{displaymath}
\Bigl[{2k\over\mu\omega_{T}}-\omega{\partial \epsilon\over\partial k}
+{k^{2}\over\omega_{T}}{\partial\over\partial k}{1\over\mu}
\Bigr]
{\partial\over\partial z}{|E_{T}|^{2}\over 2}
+\Bigl[{\partial \over\partial\omega}(\omega\epsilon)
-{\partial\over\partial \omega}\Bigl({k^{2}\over\mu\omega}\Bigr)
\Bigr]
{\partial\over\partial t}{|E_{T}|^{2}\over 2}.\end{displaymath}
It is convenient to define
$f(k,\omega)\equiv\omega\,\epsilon-k^{2}/(\omega\mu).$
Then the work-energy theorem can be stated as
\begin{equation}
-{\partial f\over\partial k}\;{\partial\over\partial z}{|E_{T}|^{2}\over 2}
+{\partial f\over\partial\omega}
\;{\partial\over\partial t}{|E_{T}|^{2}\over 2}
=-{\rm Re}(E_{x}J_{x\;\rm ext}^{*}).\label{TPoyn3}\end{equation}
The  energy flux and energy density are
\begin{equation}
T^{30}(z,t)=-{\partial f\over\partial k}\;|E_{T}|^{2}
\hskip1cm 
T^{00}(z,t)={\partial f\over\partial\omega}
\;|E_{T}|^{2}.\label{T00}\end{equation}
Relation  (\ref{mu}) allows $f$ to be expressed entirely in 
terms of the transverse
self-energy:
$f(k,\omega)=(K^{2}-\Pi_{T})/\omega$.
The $k$ and $\omega$ derivatives in (\ref{TPoyn3}) are to be 
evaluated at 
$\omega_{T}$, i.e. at $f=0$. Therefore the energy is transported
 at the group velocity of the
wave:
\begin{equation}
T^{30}(z,t)\Big/T^{00}(z,t)=-\Bigl({\partial f\over\partial k}\Big/
{\partial f\over\partial\omega}\Bigr)_{f=0}={d\omega_{T}\over dk}.
\end{equation}
Because $\partial f/\partial \omega$ is related to
 the residue $Z_{T}$ through (\ref{Ztrans}), the energy density can be
 written
\begin{equation}
 T^{00}(z,t)={2\over Z_{T}}|E_{T}|^{2}.\end{equation}

{\it Plane Wave Limit.} The final test of 
 $T^{00}$ is that it give the correct
energy density in the plane wave limit: $J^{\mu}_{\rm ext}\to 0$. The
electric field  magnitude of the plane wave TEM gluon
oscillation is fixed by the propagator to be $|E_{T}|^{2}=
\omega_{T}Z_{T}/2V$ in (\ref{EBtrans}). 
Thus the energy density of one transverse gluon is
$T^{00}=\omega_{T}/V$ as it should be.

\subsection{Momentum Conservation}

For the TEM mode the equation for momentum conservation (\ref{mom1})
 is
\begin{equation}
{\partial\over\partial t}(D_{x}^{*}B_{y})
+{\partial H_{y}^{*}\over\partial z}D_{y}
+D_{x}^{*}{\partial E_{x}\over\partial z}
=-J_{x\;\rm free}^{*}
B_{y}\end{equation}
Substituting the fields (\ref{B})-(\ref{H}) and taking the real part
gives \begin{equation}
-{k\over\omega_{T}}{\partial f\over\partial k}\;{\partial\over
\partial z}{|E_{T}|^{2}\over 2}
+{k\over\omega_{T}}{\partial f\over\partial \omega}\;{\partial
 \over\partial t}{|E_{T}|^{2}\over
2}=-{\rm Re}(J_{x\;\rm free}^{*}B_{y}).\end{equation}
The momentum flux and momentum density are therefore
\begin{equation}
T^{33}=-{k\over\omega_{T}}{\partial f\over\partial k}\;|E_{T}|^{2}
\hskip1cm
T^{03}={k\over\omega_{T}}{\partial f\over\partial \omega}\;
|E_{T}|^{2}\end{equation}
The momentum is transported at the group velocity of the
wave:
\begin{equation}
T^{33}(z,t)\Big/T^{03}(z,t)=-\Bigl({\partial f\over\partial k}\Big/
{\partial f\over\partial\omega}\Bigr)_{f=0}={d\omega_{T}\over dk}.
\end{equation}
Using (\ref{Ztrans}) one can  write the momentum density as
\begin{equation}
T^{03}={2k\over\omega_{T}Z_{T}}\;|E_{T}|^{2}\end{equation}
With the plane wave value of $E_{T}$  this gives the correct value
$T^{03}=k/V$.

\section{Longitudinal gluonic oscillations}\label{sec:Long}

\subsection{Modulated fields radiated from a current}

The longitudinal plane wave (2.22)  comes from  a pole in the propagator
 at $(F\cdot H)^{2}=0$ satisfying Maxwell's equations (\ref{2.4.3}) with
$J_{\rm free}=0$.
To identify the energy and momentum content one needs a modulated wave
\begin{equation}
E_{z}=e^{i(kz-\omega_{L} t)}\;E_{L}(z,t),\end{equation}
where the amplitude $E_{L}(z,t)$ is  slowly varying in space 
and in time.
 The only significant variation should be at $|z|\to\infty$
and $|t|\to\infty$. 
It has the Fourier representation
\begin{equation}
E_{L}(z,t)=\int d\alpha d\beta\; C_{L}(\alpha,\beta)
\;e^{i(\alpha z-\beta t)},\end{equation}
where the function $C_{L}(\alpha,\beta)$ has support only for small
 $\alpha$ and
small $\beta$. 
The electric field  has an associated
displacement field
\begin{equation}
D_{z}=e^{i(kz-\omega_{L} t)}\int d\alpha d\beta\;
\epsilon(k',\omega')C_{L}(\alpha,\beta) \;e^{i(\alpha z-\beta t)},
\label{DL}\end{equation}
where
$k'\equiv k+\alpha$ and $\omega'\equiv\omega_{L}+\beta$ as before. Both
 $\pol{H}$ and $\pol{B}$ are
identically zero. 
The modulated wave  requires 
an external gluon current with components
\begin{equation}
J_{z\,\rm free}=e^{i(kz-\omega_{L}t)}\int d\alpha d\beta
\;i\omega'\epsilon(k',\omega')C_{L}(\alpha,\beta)
\;e^{i(\alpha z-\beta t)}\end{equation}
\begin{equation}
J_{0\,\rm free}=e^{i(kz-\omega_{L}t)}\int d\alpha d\beta
\;ik'\epsilon(k',\omega')C_{L}(\alpha,\beta)
\;e^{i(\alpha z-\beta t)}\end{equation}
In the plane wave limit $C_{L}\to \delta(\alpha)\delta(\beta)$
 and therefore
$\epsilon(k',\omega')\to\epsilon(k,\omega_{L})=0$. This means that 
$D_{z}\to 0$, $J_{z}\to 0$, and $J_{0}\to 0$ eventually.

Before taking the plane wave limit,   the modulations in $D_{z}$
 are different than those of $E_{z}$  because the former includes 
the field of the polarization charges. To
approximate $D_{z}$, expand $\epsilon(k',\omega')$ 
in a Taylor
series: 
\begin{equation}
D_{z}= e^{i(kz-\omega_{L} t)}\int d\alpha d\beta
\Bigl(\alpha{\partial \epsilon\over\partial k}+\beta{\partial 
\epsilon
\over\partial\omega}+\cdots\Bigr)C_{L}(\alpha,\beta) 
\;e^{i(\alpha z-\beta
t)}.\end{equation}
 In terms of $E_{L}$ this means that
\begin{equation}
D_{z}= e^{i(kz-\omega_{L} t)}\;\Bigl(-i{\partial\epsilon\over\partial k}
{\partial E_{L}\over\partial z} +i{\partial\epsilon\over\partial \omega}
{\partial E_{L}\over\partial t}+\cdots\Bigr).\label{Dz}\end{equation}
The derivatives of $\epsilon$ are evaluated at $\epsilon=0$.

\subsection{Energy Conservation}

Since the only nonvanishing fields for the  modulated longitudinal mode
are $E_{z}$ and  $D_{z}$, the work-energy theorem
(\ref{eng1}) is
 \begin{equation}
{\rm Re}\bigl({\partial D_{z}^{*}\over\partial t}E_{z}\bigr)
=-{\rm Re}(J^{*}_{z\;\rm free}E_{z}).\end{equation}
Despite its rather trivial appearance this equation does describe the flow of energy in the
plasma.
  Because of (\ref{Dz})
\begin{equation}
{\partial D_{z}^{*}\over\partial t}E_{z}=
E_{L} \Bigl(-\omega_{L}{\partial\epsilon\over\partial k}
{\partial E_{L}^{*}\over\partial z} +\omega_{L}{\partial\epsilon
\over\partial \omega}
{\partial E_{L}^{*}\over\partial t}+\cdots\Bigr).\end{equation}
Taking the real part, with $\epsilon$ assumed real, gives
\begin{equation}
-\omega_{L}{\partial\epsilon\over\partial k}
{\partial\over\partial z}{|E_{L}|^{2}\over 2}+
\omega_{L}{\partial\epsilon\over\partial \omega}
{\partial\over\partial t}{|E_{L}|^{2}\over 2}
=-{\rm Re} \bigl(E_{z}J_{z\;\rm free}^{*}\bigr).\end{equation}
Thus the energy flux and the energy density are
\begin{equation}
T^{30}=-\omega_{L}{\partial\epsilon\over\partial k}\;
|E_{L}|^{2}\hskip1cm 
T^{00}  =\omega_{L} {\partial\epsilon\over\partial\omega}
\;|E_{L}|^{2},\label{ELong}\end{equation}
with the derivatives  evaluated at $\epsilon=0$.
The energy is transported at the group velocity of the longitudinal
 wave:
\begin{equation}
T^{30}(z,t)\Big/T^{00}(z,t)=-\Big({\partial \epsilon\over\partial k}
\Big/
{\partial\epsilon\over\partial \omega}\Big)_{\epsilon=0}=
{d\omega_{L}\over dk}\end{equation}
Because of (2.18) and (\ref{eps}) 
\begin{equation}
{\partial\epsilon\over\partial\omega}\Big|_{\epsilon=0}={1\over
f_{1}^{2}K^{2}}{2\omega_{L}\over Z_{L}}.\end{equation}
The energy density is therefore
\begin{equation}
T^{00}={2\omega_{L}^{2}\over f_{1}^{2}K_{L}^{2}\,Z_{L}}\;|E_{L}|^{2}\end{equation}

{\it Plane Wave Limit.} Whenever the external current is
 present the space
and time derivatives of $|E_{L}(z,t)|^{2}$ prescribe the energy 
transport within the plasma.
 When the external
current is removed, the value of $T^{00}$ remains constant. The
 value of that
constant is the energy contained in the wave.
The pole in the propagator fixes the amplitude for the
free longitudinal gluon mode to be  $|E_{L}|^{2}=f_{1}^{2}Z_{L}\,K_{L}^{2}
/(2\omega_{L}V)$ as shown in
(\ref{Elong}).  The  energy density of this mode is
therefore $ T^{00}=\omega_{L}/V$.

\subsection{Momentum conservation}

For the longitudinal mode the force equation (\ref{mom1}) is
\begin{equation}
{\partial D^{*}_{z}\over\partial z}\,E_{z}=-J^{0*}_{\rm free}
\,E_{z}.\end{equation}
Note that although $J_{z\;\rm free}$ is non-zero, it exerts no
 force on the plasma charges in this
mode because $\pol{B}=0$. Substituting (\ref{Dz}) and taking the 
real part  gives
\begin{equation}
-k{\partial \epsilon\over\partial k}\,{\partial
 \over\partial z}{|E_{L}|^{2}\over 2}
+k{\partial \epsilon\over\partial \omega}\,{\partial
 \over\partial t}{|E_{L}|^{2}\over 2}
=-{\rm Re}\big(J^{0*}_{\rm free}\,E_{z}\big).\end{equation}
Thus the momentum flux and momentum density are
\begin{equation}
T^{33}=-k{\partial \epsilon\over\partial k}\,|E_{L}|^{2}
\hskip1cm
T^{03}=k{\partial \epsilon\over\partial \omega}\,|E_{L}|^{2}.\end{equation}
Momentum is transported at the group velocity of the wave
\begin{equation}
T^{33}(z,t)\Big/T^{03}(z,t)=-\Big({\partial \epsilon\over
\partial k}\Big/
{\partial\epsilon\over\partial \omega}\Big)_{\epsilon=0}=
{d\omega_{L}\over dk}.\end{equation}
The momentum density can also be written
\begin{equation}
T^{03}={2k\omega_{L}\over f_{1}^{2}K_{L}^{2}\,Z_{L}}|E_{L}|^{2}\end{equation}
In the plane wave limit
$|E_{L}|^{2}=f_{1}^{2}Z_{L}\,K_{L}^{2}/(2\omega_{L}V)$ so that
$T^{03}=k/V$, which is the correct result.

\section{Conclusion}

Regardless of the functional dependence of $\omega_{T}$ and  $\omega_{L}$
on momentum, temperature, and quark masses the energy density carried by the field of a single
gluon  of either type has been shown to be  $\omega_{T}/V$ and  $\omega_{L}/V$, respectively.

\acknowledgments

This work was supported in part by National Science Foundation grant
 PHY-9630149.


\references

\bibitem{BP1} R.D. Pisarski, Phys. Rev. Lett. {\bf 63} (1989) 1123;
 E. Braaten and R.D. Pisarski,
Phys. Rev. Lett. {\bf 64} (1990) 1338.

\bibitem{ward1} E. Braaten and R.D. Pisarski, Nucl. Phys. {\bf B337} 
(1990) 569; {\bf B339} (1990) 310.

\bibitem{act1} E. Braaten and R.D. Pisarski, Phys. Rev. {\bf D45} (1992)
1827.

\bibitem{ward} J. Frenkel and J.C. Taylor, Nucl. Phys. {\bf B334} (1990)
199.

\bibitem{act2} J.C. Taylor and S.M.H. Wong, Nucl. Phys. {\bf B346} (1990)
115.

\bibitem{act3} J. Frenkel and J.C. Taylor, Nucl. Phys.  {\bf B374} (1992)
156.

\bibitem{act4} F.T. Brandt, J. Frenkel, J.C. Taylor, and S.M.H. Wong,
Canadian J. Phys. {\bf 71} (1993) 219. 

\bibitem{act5} J.P. Blaizot and E. Iancu, Nucl. Phys. {\bf B390} (1993)
589.

\bibitem{act6} J.P. Blaizot and E. Iancu, Phys. Rev. Lett. {\bf 70}
 (1993) 3376;  Nucl. Phys. {\bf B417} (1994) 608.

\bibitem{act7} R. Efraty and V.P. Nair, Phys. Rev. Lett. {\bf 68}
 (1992) 2891; Phys. Rev. {\bf D47} (1993) 5601.

\bibitem{act8} R. Jackiw and V.P. Nair, Phys. Rev. {\bf D48} (1993) 4991; 
 R. Jackiw, Q. Liu, and C. Lucchesi, Phys. Rev. {\bf D49} (1994) 6787.

\bibitem{em1}  H.A. Weldon, Can. J. Phys. {\bf 71} (1993) 300.

\bibitem{em2} F.T. Brandt, J. Frenkel, and J.C. Taylor, Nucl. Phys.
 {\bf B410} (1993) 3; erratum, Nucl. Phys. {\bf B419} (1994) 406;

\bibitem{em3} V.P. Nair, Phys. Rev. {\bf D48} (1993) 3432.

\bibitem{em4} J.P. Blaizot and E. Iancu, Nucl. Phys. {\bf B421} (1994) 565.

\bibitem{em5}  F.T. Brandt, J. Frenkel, and J.C. Taylor,
Nucl. Phys. {\bf B437} (1995) 433; J. Frenkel, E.A. Gaffney, and J.C.
Taylor, Nucl. Phys. {\bf B439} (1995) 131; E. Gaffney, Nucl. Phys. 
{\bf B442} (1995) 268.

\bibitem{AW3} H.A. Weldon, ``Structure of the Gluon Propagator at Finite
Temperature", to appear in Ann. Phys. (N.Y.),  hep-ph/9701279.

\bibitem{KKR} R. Kobes, G. Kunstatter, and A. Rebhan, Phys. Rev. Lett.
{\bf 64} (1990) 2992 and Nucl. Phys. {\bf B335} (1991) 1. 


\bibitem{Br} L. Brillouin, {\it Wave Propagation and Group Velocity}
 (Academic Press, New York, 1960) 88-95.

\bibitem{Ginz} V.L. Ginzburg, {\it The Propagation of Electromagnetic
 Waves in Plasmas} (Pergamon
Press, New York, 1970) 293-296.

\bibitem{LL} L.D. Landau, E.M. Lifshitz, and L.P. Pitaevskii, 
{\it Electrodynamics of Continuous
Media} (Pergamon Press, New York, 1984) 272-276.

\bibitem{22} U. Heinz, K. Kajantie, and T. Toimela,
Ann. Phys. (N.Y.) {\bf 176} (1987) 218.

\bibitem{23} H.T. Elze, U. Heinz, K. Kajantie, and T. Toimela,
Z. Phys. C {\bf 37} (1988) 305.

\bibitem{24} R. Kobes, G. Kunstater, and K.W. Mak, Z. Phys. C
{\bf 45} (1989) 129.

\bibitem{25} F. Flechsig and H. Schulz, Phys. Lett. {\bf B349}
(1995) 504. 

\bibitem{26} D.J. Gross, R.D. Pisarski, and L.G. Yaffe,
Rev. Mod. Phys. {\bf 53} (1981) 43.

\bibitem{27} K. Kajantie and J. Kapusta, Ann. Phys. (N.Y.)
{\bf 160} (1985) 477.

\bibitem{28} G. Kunstatter, Can. J. Phys. {\bf 71} (1993) 256.
\end{document}